\documentclass{article}
\usepackage{arxiv}
\usepackage[utf8]{inputenc}
\usepackage{amsmath,amssymb,amsfonts}
\usepackage{algorithmic}
\usepackage{graphicx}
\usepackage{textcomp}
\usepackage{xcolor}
\usepackage{booktabs}
\usepackage{makecell}
\usepackage{balance}
\usepackage[nolist]{acronym}
\usepackage[numbers]{natbib}

\usepackage{bm}
\usepackage{moresize}
\usepackage{url}
\usepackage{tikz}
\usepackage[most]{tcolorbox}
\usepackage{graphicx}
\usepackage{subfig}
\usepackage{enumitem}
\usepackage{xcolor,colortbl}
\definecolor{Gray}{gray}{0.85}

\newcolumntype{C}{>{\centering\let\newline\\\arraybackslash\hspace{0pt}}m{0.15\textwidth}}
\newcolumntype{V}{>{\centering\let\newline\\\arraybackslash\hspace{0pt}}m{0.4\textwidth}}
\newcolumntype{L}{>{\let\newline\\\arraybackslash\hspace{0pt}}m{0.4\textwidth}}
\newcolumntype{M}{>{\let\newline\\\arraybackslash\hspace{0pt}}m{0.1\textwidth}}

\begin{document}

\newcommand{\trtitle}{Continual Learning for Affective Robotics: A Proof of Concept for Wellbeing}
\newcommand{\nl}{\vspace{0.5mm}{\noindent}}

\title{\trtitle}

\author{
Nikhil~Churamani$^{*,\dagger}$, Minja~Axelsson$^{*,\dagger}$, Atahan~Çaldır$^{\ddagger}$ and Hatice Gunes$^{\dagger}$\\
$^\dagger${Department of Computer Science and Technology, University of Cambridge, United Kingdom.}\\
\texttt{Email: \{nikhil.churamani, minja.axelsson, hatice.gunes\}@cl.cam.ac.uk}\\
$^{\ddagger}${Department of Computer Science, Özyeğin University, Istanbul, Turkey} \\
\texttt{Email: atahan.caldir@ozu.edu.tr}\thanks{Equal Contribution.}
}

\maketitle

\begin{abstract}
Sustaining real-world human-robot interactions requires robots to be sensitive to human behavioural idiosyncrasies and adapt their perception and behaviour models to cater to these individual preferences. For affective robots, this entails learning to adapt to individual affective behaviour to offer a personalised interaction experience to each individual. \acf{CL} has been shown to enable real-time adaptation in agents, allowing them to learn with incrementally acquired data while preserving past knowledge. In this work, we present a novel framework for real-world application of \ac{CL} for modelling personalised human-robot interactions using a \ac{CL}-based affect perception mechanism. To evaluate the proposed framework, we undertake a \textit{proof-of-concept} user study with $20$ participants interacting with the Pepper robot using three variants of interaction behaviour: static and scripted, using affect-based adaptation without personalisation, and using affect-based adaptation with \textit{continual personalisation}. Our results demonstrate a clear preference in the participants for \ac{CL}-based \textit{continual personalisation} with significant improvements observed in the robot's \textit{anthropomorphism, animacy} and \textit{likeability} ratings as well as the interactions being rated significantly higher for \textit{warmth} and \textit{comfort} as the robot is rated as significantly better at understanding how the participants \textit{feel}. 

\nl \textbf{Keywords:} Continual Learning, Affective Robotics, Wellbeing, Facial Affect, Human-Robot Interaction

\end{abstract}

\section{Introduction and Related Work}

Social and affective robots are designed to interact in human-centred environments, learning to sustain closed-loop interactions with humans. Acting as assistants~\cite{Leite2013SocialRF}, tutors or coaches~\cite{Bodala2021Teleoperated,GASKESW18}, and companions that offer personalised conversational capabilities~\cite{Rudoviceaao6760}, they need to sense and understand human behaviour and learn to support them with context-appropriate social interactions, aiding and even fostering their cognitive and socio-emotional wellbeing~\cite{Dautenhahn2004Robots}. This entails socio-emotional adaptability in affective robots that may allow them to not only adapt to real-world interactions with individuals but also provide them personalised interaction experiences by adapting to individual context and behaviour~\cite{ayub2020b, Churamani2020CLIFER,Rudoviceaao6760}. Developing sensitivity towards specific verbal and non-verbal human behaviours~\cite{Churamani2020CLAC,Churamani2020CL4AR} while learning to respond dynamically and appropriately enables social robots to offer personalised and naturalistic human-robot interactions~\cite{Ficocelli2016Promoting}.

Humans use affect to convey meaning and intent in conversations using several outward signals~\cite{Gunes2011Emotion} such as facial expressions~\cite{Li2020Deep}, body gestures~\cite{Noroozi2021Survey} or speech intonations~\cite{Schuller2018SER}. For affective robots, it is essential to interpret such affective signalling to gather a contextual understanding of their interactions with humans. This may enable them to adapt their own behaviours during interactions to appropriately reflect the users’ affective state, engaging them in meaningful conversations~\cite{Churamani2022Affect, Ficocelli2016Promoting}. However, this may not be easy in all situations owing to the unpredictability of human-centred environments resulting from changing environmental conditions and interaction contexts and the inherent uncertain and subjective nature of human affective expression~\cite{Dziergwa2018}. 

Affective Robotics research explores how embedding robots with an understanding of human affective behaviour may allow them to facilitate meaningful interactions that enhance an individual's experience with the robot~\cite{KIRBY2010Affective}. Yet, most approaches make use of off-the-shelf affect perception models that, despite providing state-of-the-art results on benchmarks, are not able to adapt to the dynamics of real-time interactions. Any adaptation in the model requires a large amount of computational resources as they need to be trained, from scratch in most cases, with large amounts of data. This may not be feasible, especially when applied on resource-constrained devices such as robots in real-world situations~\cite{Churamani2020CL4AR,dlAtTheEdge}. 

\acf{CL}~\cite{LESORT2020CL4R, Parisi2018b} for robots focuses, instead, on adapting to changing contexts and shifts in the inherent data distributions. It aims to instil long-term adaptability in agents integrating novel information, acquired incrementally by the agent as it interacts with the environment, while balancing past knowledge~\cite{LESORT2020CL4R}. For affective robots, this entails adapting and personalising to individual socio-emotional behaviours over repeated interactions with the users, learning to adapt their own behaviours in the process. The desiderata for \textit{continually adapting} affective robots thus becomes to model personalisation capabilities that allow robots to be sensitive to individual differences in affective behaviours~\cite{Churamani2020CLIFER, pmlrbarros19a} and enable them to model context-appropriate behaviours during interactions~\cite{Churamani2020CL4AR, Hemminghaus2017Adaptive, McQuillin2022RoboWaiter}. 

In our previous work~\cite{Churamani2020CL4AR}, we provide a theoretical framework for \ac{CL} for affective robotics, proposing two specific components of such interactions; \textit{personalised affect perception} and \textit{context-appropriate behavioural learning}, that allow robots to  \textit{continually} learn and adapt in real-world situations. Enabling personalised affect perception requires robots to adapt their perception towards individual users, accounting for individual differences in expression and other characteristic attributes~\cite{Churamani2020CLIFER}. Several approaches achieve this by considering contextual attributes such as gender and culture~\cite{Rudoviceaao6760} to learn person-specific feature representations~\cite{Chu2017Selective, Rudoviceaao6760}, or learning individualised \textit{affective memories}~\cite{pmlrbarros19a} that encode the robot's affective experience interacting with an individual, or simultaneously learning personalised (forming an episodic memory) and generalised (forming a semantic memory) representations~\cite{Churamani2020CLIFER} that may allow robots to adapt towards each individual while preserving a generalised understanding of human affective expression across individuals and contexts. Furthermore, using such personalised models to encode the users' affective state as a contextual affordance, robots may dynamically model and adapt their interactions with the users instead of following the same static and scripted interaction for each user~\cite{Churamani2017TheImpact, Hemminghaus2017Adaptive}. Such personalised interactions are expected to enhance the users' experience interacting with the robot, offering a naturalistic interaction experience.

Extending our theoretical formulations presented in~\cite{Churamani2020CL4AR}, in this work, we present a practical framework for real-world application of continual learning for enabling personalised \ac{HRI} for affective robots. The presented framework enables a multi-modal interaction, with \textit{speech recognition} capabilities to parse user responses, and a \ac{CL}-based \textit{facial affect perception} model~\cite{Churamani2020CLIFER} that personalises its learning towards each individual's expressions. A \textit{dialogue manager} allows to structure the interactions into different interaction states based on the interaction context. The robot uses its evaluation of participants' affective behaviour to model state-transitions, personalising interactions to individual preferences and employs \textit{\ac{NLG}} to generate naturalistic responses towards the participants. To evaluate the contribution of \ac{CL}-based personalisation towards how the participants' impressions of the robot, we implement another version of the framework, for comparison, switching the \ac{CL}-based model with the \textit{FaceChannel}~\cite{BCS2020FaceSN, barros2020facechannel}, an off-the-shelf state-of-the-art facial affect perception model that still encodes the participants' facial affect, albeit without any personalisation. Furthermore, a control condition is implemented by completely `switching off' affective adaptation and always following a static and scripted interaction flow, ignoring the participants' affective responses. For the \textit{proof-of-concept} user study, $20$ participants interact with three versions of the Pepper robot, in a \textit{between-subjects} design; (i)~following a \textit{static and non-adaptive} interaction script, (ii)~using affective adaptation \textit{without} personalisation, or (iii)~using \ac{CL}-based affective adaptation \textit{with} personalisation. Our results demonstrate that using \ac{CL}-based personalisation improves participants' impressions of the robot for \textit{anthropomorphism, animacy} and \textit{likeability} while offering \textit{warm} and \textit{comfortable} interactions by being sensitive to the participants \textit{feelings}.

\section{The Interaction Scenario}
\label{sec:scenario}

Enabling personalised interaction capabilities in affective robots can be beneficial for \ac{HRI} across a variety of interaction contexts. In particular, we explore sensing and adapting to human affective behaviour in a one-off \ac{PP}-based~\cite{cunha2019positive,lopez2018positive} interaction session where the Pepper Robot\footnote{\scriptsize\url{https://www.softbankrobotics.com/emea/en/pepper}} requests participants to reflect upon recent events in their lives that may invoke positive feelings in them. 

We develop a script for the interaction scenario in collaboration with a professional psychologist to mitigate situations or utterances from the robot that may risk invalidating or patronising the participants. This also allows us to improve the interaction, clarifying prompts for participants to share their experiences, and determine the appropriate amount of robot utterances. Participant responses are modelled in two ways; `Yes/No' questions determining the flow of the interactions, and open-ended `descriptive' questions where the participants talk about their experiences in as much detail as they deem appropriate. Participants' affective behaviour during the descriptive questions is observed and used to select appropriate robot responses using empathetic utterances~\cite{leite2013influence} that are reflective of the participants' affective state. Each interaction with the participants consists of three exercises or tasks:

\begin{enumerate}
    \item The robot asks participants to talk about $2$ impactful things or events in their lives from the past two weeks, why these events may have happened and how the events made them feel. 
    \item Focusing on developing \emph{gratitude} to increase positive affect and subjective happiness, the robot asks the participant to recall $2$ things that they felt grateful for in the recent past. 
    \item The robot then asks the participants to describe $2$ recent \emph{accomplishments}, the strengths applied to accomplish these, and how these made them feel. 
\end{enumerate}

\noindent After the interaction, the robot asks participants for verbal and survey-based feedback using a tablet.


\section{The Proposed Framework}

\begin{figure}
    \centering
    \includegraphics[width=\textwidth]{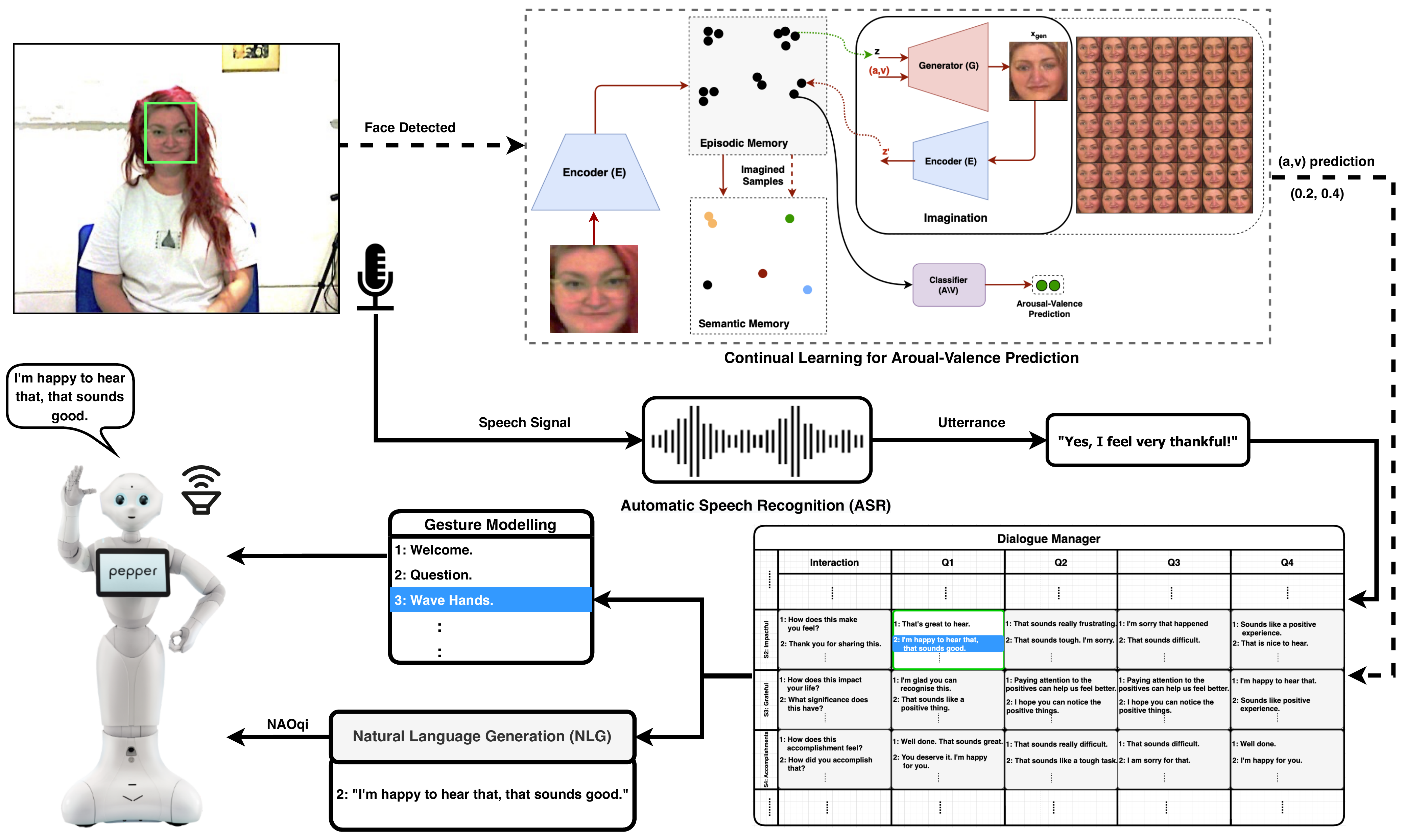}
    \caption{The Proposed Framework: Continual Learning for Personalised Human-Robot Interactions.
    }
    \label{fig:framework}
\end{figure}

To facilitate interactions with Pepper, we implement a modular framework that enables multi-modal interaction capabilities in the robot (see Fig.~\ref{fig:framework}). This allows us to selectively employ different functionalities as needed to implement different variations in robot behaviours. All modules are implemented on Pepper as \ac{ROS}\footnote{\url{http://wiki.ros.org/melodic}} modules communicating with each other.

\subsection{Face Detection and Obtaining Ground Truth}
We use Pepper's on-board RGB-camera (from the forehead) to record participants' behaviour at $30$ FPS with a resolution of $640\times480$. OpenCV Face Detection\footnote{\scriptsize\url{https://github.com/opencv/opencv}} is used to detect and crop-out the participants' face from the recorded images. These face images are passed to the \textit{FaceChannel}~\cite{BCS2020FaceSN, barros2020facechannel} to annotate the participants' real-time facial expressions in terms of \textit{valence}, representing the positive or negative nature of their expression and \textit{arousal}, representing the intensity, normalised to $\in[-1,1]$, providing \textit{ground truth} evaluations. Each annotated frame provides \textit{labelled} data to be used by the \ac{CL}-based personalisation model. These annotated frames are published as \ac{ROS} parameters to be processed by the rest of the system. For \textit{affect-based adaptation without personalisation}, annotations over the last $5$ seconds ($5\times30=150$ frames) of each dialogue provide an evaluation of the participants' affective responses towards the robot and are used to modulate the interaction flow (see Section~\ref{subsec:ISM}).

\begin{figure}
    \centering
    \includegraphics[width=0.6\textwidth]{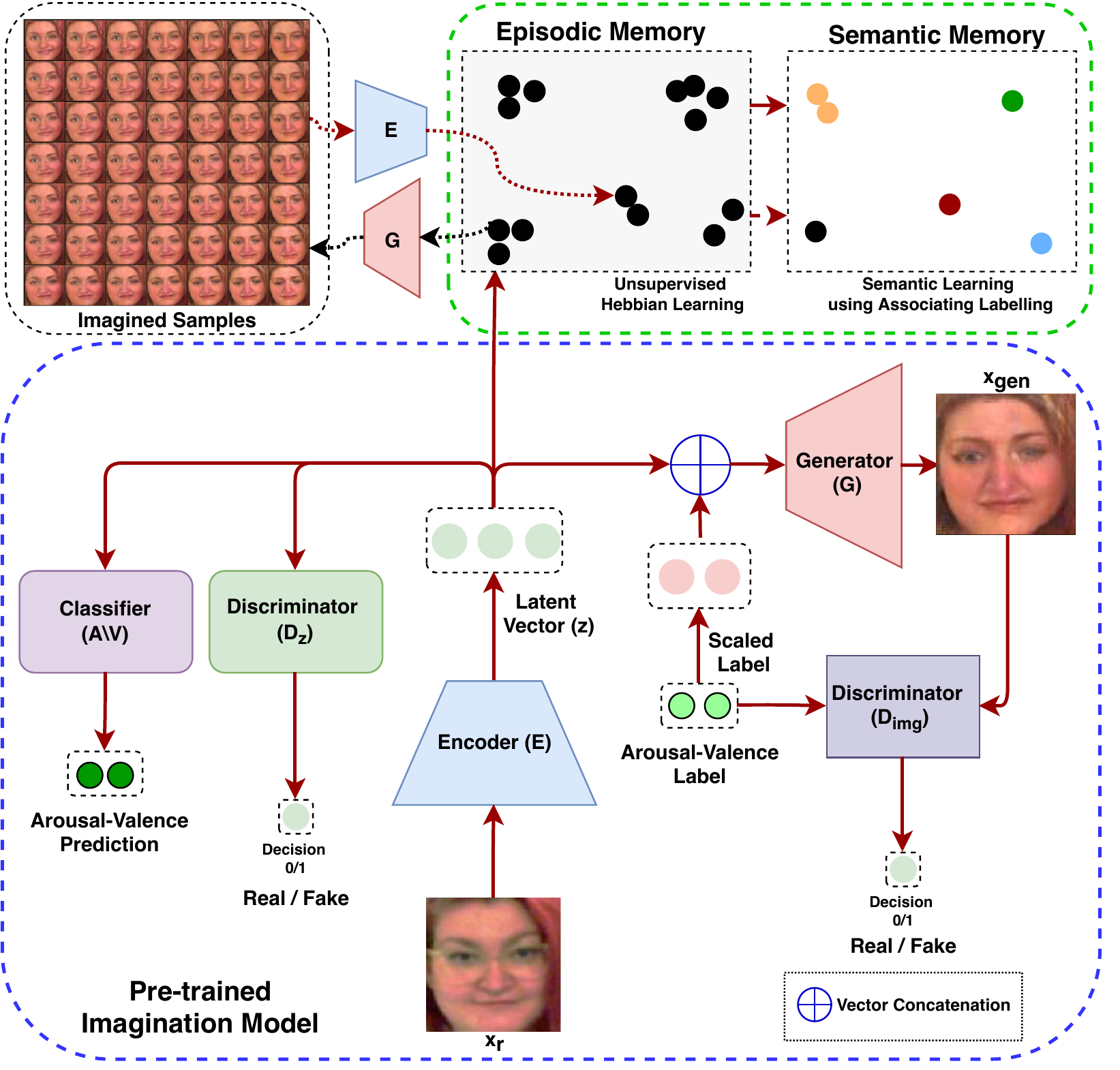}
    \caption{Adapting \acs{CLIFER}~\cite{Churamani2020CLIFER} for \textit{Imagining} participant faces for a range of arousal-valence values.}
    \label{fig:imagination}
\end{figure}

\subsection{Continual Learning for Personalised Affect Perception}
\label{sec:clifer}
For \textit{personalised affect perception} we use the \acs{CLIFER} framework~\cite{Churamani2020CLIFER}, initially evaluated only on benchmark datasets, by extending its \textit{imagination} model for dimensional facial expression editing~\cite{Lindt2019FG} and applying it in the context of \ac{HRI}. \acs{CLIFER} implements a \acf{GDM}-based architecture~\cite{Parisi2018a} that allows for incrementally personalising the robot's perception model towards an individual. It consists of an \textit{episodic} (\acs{GDM}-E) and a \textit{semantic} (\acs{GDM}-S) memory, both of which use a \acf{GWR} neural network with Gamma-filtering~\cite{Parisi2018a} for a semi-supervised processing of facial features for affect perception~\cite{Churamani2020CLIFER}. While the \textit{episodic} memory learns instance-level variations, personalising to the participants' current expressions during the interactions, the \textit{semantic} memory learns concept-level abstractions for long-term retention of knowledge over repeated interactions with a participant. Imagination in \acs{CLIFER} allows for the simulation of additional facial images for a participant by conditionally translating their face to represent different affective annotations using a \ac{CAAE}-based adversarial learning model~\cite{Churamani2020CLIFER, Lindt2019FG}. These generated images augment learning in the dual-memory \ac{GDM} model that personalises towards each individuals' expressions.

During the interactions, from each response by the participant, we randomly sample $10$ \textit{FaceChannel}-annotated images of the participant, and for each of these images, \acs{CLIFER} \textit{imagines} or simulates $49$ additional facial images (see Fig.~\ref{fig:imagination}), corresponding to \textit{arousal} and \textit{valence} values ranging from ($-0.75$ to $0.75$) (values capped at $||0.75||$ to avoid extremes). The additional $10\times49 = 490$ images, along with the original $10$ images, are then passed through the encoder to generate feature representations that are used to update the \acs{GDM}-based learning model. As the robot interacts with the participant, it is able to simulate \textit{imagined contact}~\cite{Wullenkord2019} over a wide-range of affective contexts, personalising towards their expressions. For real-time affect perception, we use the \textit{episodic} memory representations to summarise the arousal-valence predictions over the last $5$ seconds ($5\times30=150$ frames) of each dialogue and publish them as \ac{ROS} parameters to be used by the dialogue manager to modulate the interaction flow.

\subsection{Dialogue Management}
\subsubsection{Interaction State Manager}
\label{subsec:ISM}
We model a bi-directional interaction where Pepper listens to the participants' and responds appropriately by generating naturalistic and congruous responses. For this, we implement a \ac{FSM}-based dialogue manager using the SMACH\footnote{\scriptsize\url{http://wiki.ros.org/smach}} \ac{ROS} library. The interactions consist of $7$ different dialogue states, starting with \textit{Introduction} ($S_1$) followed by the three interaction states tagged as \textit{Impactful} ($S_2$), \textit{Grateful} ($S_3$) and \textit{Accomplishments} ($S_4$) (see Section~\ref{sec:scenario}). After the three interactions, the participants provide a verbal \textit{Feedback} ($S_5$), fill out the \textit{Survey} ($S_6$) questionnaires and finally end the session by saying \textit{GoodBye} ($S_7$).

Under each of the interaction states ($S_2$, $S_3$ and $S_4$), participants respond to open-ended descriptive questions about their recent experiences. The robot encodes their facial affect using arousal-valence predictions from the \acs{CLIFER} (or the \textit{FaceChannel}, depending on the condition) model and divides the dialogue states further, mapping them to the four quadrants (\textit{Q$_1$}: positive valence \& positive arousal, \textit{Q$_2$}: negative valence \& positive arousal, \textit{Q$_3$}: negative valence \& negative arousal, and \textit{Q$_4$}: positive valence \& negative arousal) of the Circumplex Model of affect~\cite{russell1980circumplex}, to determine robot responses. An additional \textit{Neutral} dialogue-state is mapped for arousal-valence values $\in[-0.10, 0.10]$. These sub-states extend the dialogue with the robot adapting its responses based on the participants' affective behaviour. State transitions are adapted to generate appropriate robot responses by traversing these sub-states for each interaction state. For instance, if a participant expresses positive arousal and valence (Q$_1$) while talking about \textit{impactful} events (S$_2$), the robot utters an additional positive response such as, ``\textit{That sounds great, I'm happy for you.}'' to express its understanding of their affective state, before returning to the interaction script ($S2\xrightarrow{Q1}S3$). For the static and scripted interactions, such affective dialogue is not held and the robot strictly follows the interaction script, transiting from one interaction state to the other directly ($S2\xrightarrow{}S3$).

\begin{figure}
    \centering
    \includegraphics[width=0.5\textwidth]{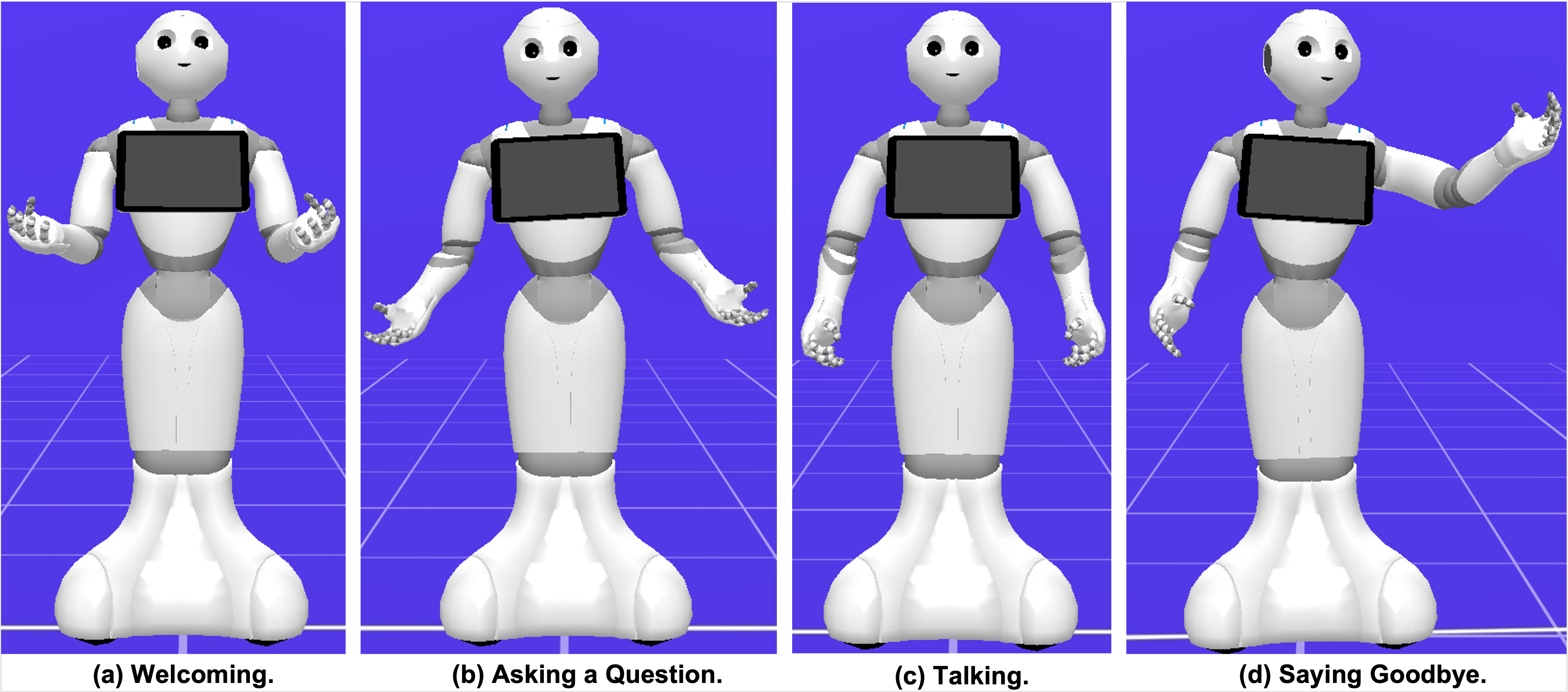}
    \caption{Pepper displaying gestures during the interactions.}
    \label{fig:gestures}
\end{figure}

\subsubsection{Speech Recognition}
\label{sec:speech}
To capture participant responses during the interactions, we use the Python Speech Recognition Library\footnote{\scriptsize\url{https://github.com/Uberi/speech_recognition}}, built on top of Google \acs{ASR}. For the `Yes/No' questions, we implement \textit{keyword-spotting} to capture variations of affirmative (\textit{yes, yeah, yep, OK, fine, aye, definitely, certainly, exactly, of course, positive, sure}) and negative (\textit{no, nope, na, never, nah, nay}) user responses. These responses are used by the state manager to model state-transitions between dialogue states ($S_1-S_7$). Other user responses to the open-ended \textit{descriptive} questions are also captured and logged but do not affect the interaction flow.

\subsubsection{\acf{NLG}}
For each interaction state in the dialogue manager, pre-defined sentence dictionaries are used by the \ac{NLG} module to generate corresponding robot responses. The dialogue manager is tasked with determining the interaction state (and the affective sub-state, depending on the condition) and selecting the sentence to be uttered by the robot. A total of $120+$ robot responses are scripted, split into several sentence dictionaries, each mapped to a given dialogue state (see Fig.~\ref{fig:framework} for some examples). Pepper's in-built \ac{TTS} Python library\footnote{\scriptsize\url{http://doc.aldebaran.com/2-5/naoqi/audio/altexttospeech.html}} is used to generate robot responses.

\subsection{Gesture Modelling}
During the interactions, Pepper is designed to generate certain upper-body gestures by manipulating its joints (head, shoulders, elbows, wrists and hands) to make the interactions seem more naturalistic~\cite{Lohan2016, Miller2016}. These gestures are performed as the robot welcomes the participants, asks a question, responds to their affective expressions (only for the adaptive interactions), and at the end of the experiment to say \textit{goodbye}. All gestures (see Fig.~\ref{fig:gestures} for examples) are pre-defined in Choregraphe\footnote{\scriptsize\url{http://doc.aldebaran.com/2-4/software/choregraphe/index.html}} after consulting the psychologist to mitigate negative feelings in the participants.

\section{Proof-of-Concept: User Study}

\subsection{Set-up}
\begin{figure}
    \centering
    \includegraphics[width=0.55\textwidth]{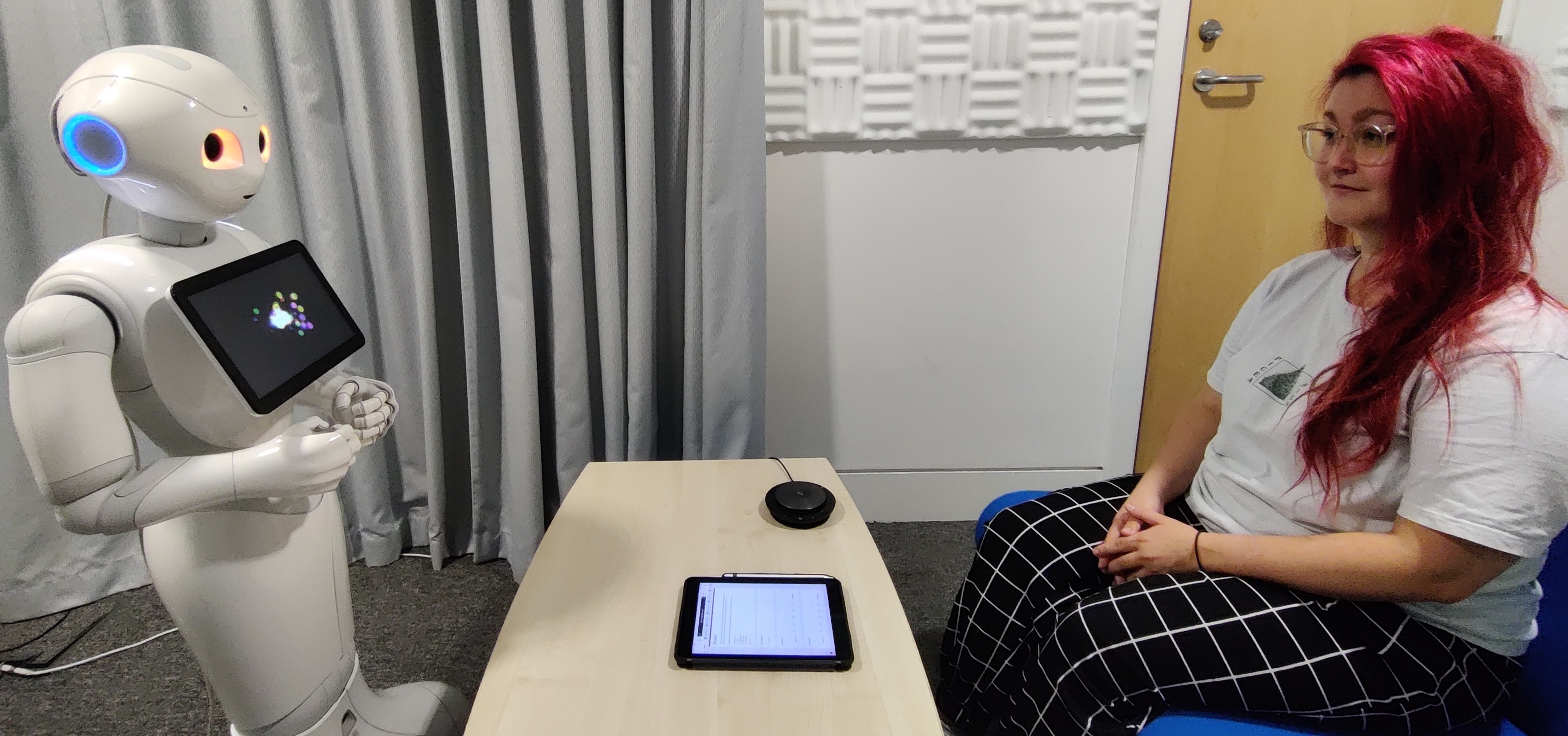}
    \caption{Setup: Pepper interacting with the Participant.}
    \label{fig:exp_setup}
\end{figure}

The participants are sat in front of Pepper with a low table separating the two, in order to keep the participants' personal space (see Fig.~\ref{fig:exp_setup}). Two cameras capture the interaction, placed facing the participant and Pepper, respectively, while Pepper's on-board RGB camera is used to capture the participants' facial affect. An external microphone is placed on the table in front of the participant to accurately record their voice. Pepper's on-board speakers are used to communicate with the participants. Additionally, a tablet is placed on the table for the participants to fill out the survey questionnaires.

\subsection{Participants}
A total of $22$ participants were recruited for the user study amongst the students and members of the university. However, some technical issues during the interaction led to the data from $2$ participants being dropped from the analyses. Thus, the evaluations presented here consist of data from $N=20$ participants ($12$ female, $5$ male, $3$ not disclosed) aged $26.70\pm3.68$ years from $12$ different nationalities. The majority of the participants ($N=13$) indicated little to no prior experience with humanoid robots. As the interaction was modelled around wellbeing and \ac{PP} exercises, only a \textit{non-clinical} population was sought making sure that none of the participants were undertaking any mental health treatment or medication. All participants were asked to fill in the \ac{GAD7}~\cite{spitzer2006brief} and \ac{PHQ9}~\cite{kroenke2001phq} before they were enrolled in the experiments to make sure no participants were experiencing high anxiety or depression. This screening resulted in several participants not being able to participate in the study creating an imbalance in the gender distribution as well as the condition groups. Before the study, participants were also asked to watch $2$ videos of Pepper showing its interaction capabilities, in order to familiarise them with the robot. All participants provided \textit{informed consent} for their participation and the usage of their data for post-study analyses. The participants were compensated in the form of Amazon vouchers. The consent form, study-design and the experimental protocol was approved by the Departmental Ethics Committee.

\subsection{Experiment Conditions}
\label{sec:conditions}
Following a \textit{between-subjects} design, each participant is randomly allocated to one of the three conditions:

\nl \textbf{C1 - Static and Scripted Interaction:} In this condition, the interaction between the participant and the robot follows the pre-defined script where the robot's responses do not take into account the participants' affective responses and the robot always responds in an anodyne manner. A total of $5$ participants were allocated to this condition.
    
\nl \textbf{C2 - Affect-based Adaptation without Personalisation:} To evaluate if embedding adaptation in the robot influences its interactions with the participants, in this condition, we employ the \textit{FaceChannel} affect perception model, to determine participants' affective expressions. Starting from the same initial state in the dialogue manager as C$1$, the affect perception outcome is used to modulate robot responses in the interaction states ($S_2, S_3$ and $S_4$) to adapt to the interaction. A total of $9$ participants were allocated to this condition.

\nl \textbf{C3 - Affect-based Adaptation with Continual Personalisation:} In this condition, instead of the pre-trained \textit{FaceChannel} model, a \ac{CL}-based \acs{CLIFER} model is employed to determine the affective state expressed by the participant during the interactions (see Section~\ref{sec:clifer} for details). As the interaction progresses, the model \textit{continually} personalises towards the individual participant's facial expressions. This personalised affect prediction is used to modulate the robot's responses towards each participant. A total of $6$ participants were allocated to this condition.

\subsection{Questionnaires}
To evaluate interactions with the robot, the participants completed a three-part questionnaire consisting of questions from the GODSPEED~\cite{bartneck2009measurement} (measuring robot \textit{anthropomorphism, animacy, likeability, perceived intelligence} and \textit{perceived safety}) and \acs{RoSAS}~\cite{carpinella2017robotic} (measuring people's perceptions of the robot for \textit{warmth}, \textit{competence}, and \textit{(dis) comfort}) questionnaires, along with customised questions on whether Pepper understood what they \textit{said} or how they \textit{felt} and \textit{adapted} its behaviour accordingly. 

\begin{figure}  
    \centering
    \subfloat[GODSPEED Scores.\label{fig:godspeed}]{\includegraphics[width=0.45\textwidth]{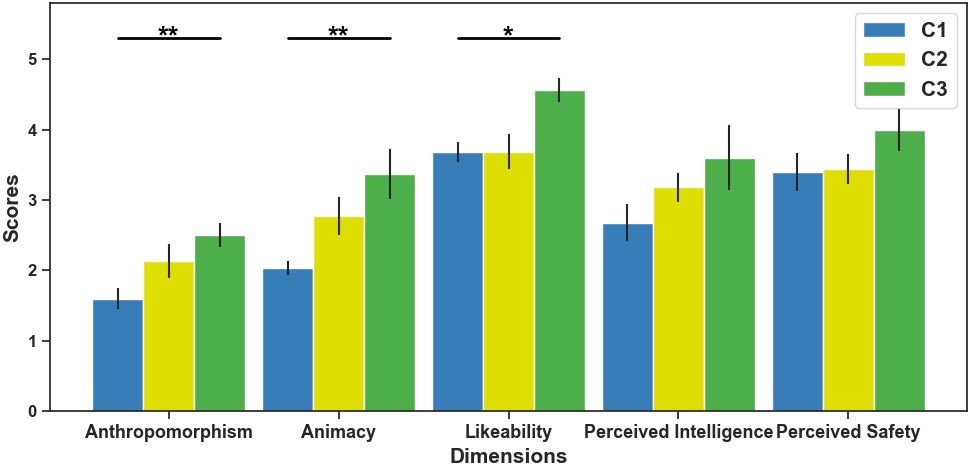}} \\\vspace{-4mm}
    \subfloat[Individual GODSPEED Dimensions.\label{fig:godspeed_dims}]{\includegraphics[width=\textwidth]{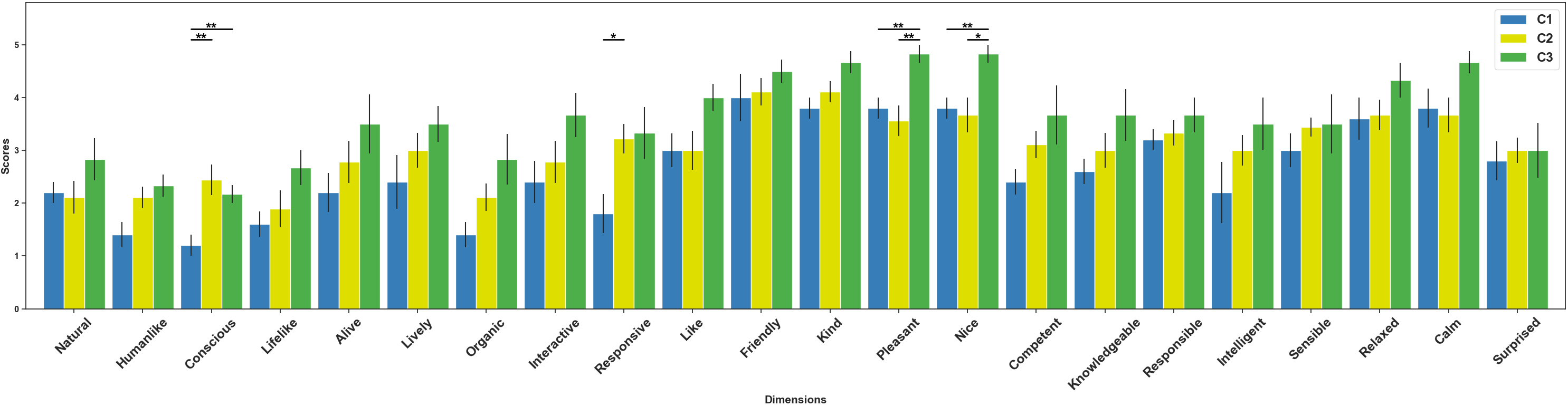}}\\
    \caption{GODSPEED~\cite{bartneck2009measurement} Scores for C$1$, C$2$, and C$3$ conditions.~$^*$ represents $p<0.05$ and $^{**}$ represents $p<0.01$.}
    \label{fig:godspeed_combined} 
    
\end{figure}
\subsection{Results}
To evaluate whether affective adaptation with \textit{continual personalisation} enhances the participants' interactions with Pepper, we examine how they rate the robot during the interactions, compared to affective adaptation without personalisation as well as the control condition following a static interaction script. Following a \textit{between-subjects} design, we compare the participants' evaluation of the robot under the three conditions. First, the three conditions are compared using the Kruskal-Wallis H Test~\cite{Kruskal1952KW} to measure significant differences between the three conditions in terms of the participants' ratings for the robot. For the the dimensions with significant differences, as a post-hoc analysis, pair-wise \textit{non-parametric} one-tailed Mann-Whitney~U Test~\cite{Mann1947} is conducted comparing C$1$~vs.~C$2$, C$2$~vs.~C$3$, and C$1$ vs. C$3$. Bonferroni correction is applied to all the statistical comparisons conducted.  Here we present the main outcomes of these evaluations:

\subsubsection{GODSPEED} 
Our results show that both C$2$ and C$3$ are, on average, rated higher than C$1$ across all the GODSPEED evaluations (see Fig.~\ref{fig:godspeed_combined}) indicating a clear preference with the participants towards an adaptive Pepper that is sensitive to their affective expressions. Furthermore, we see that \textit{continual personalisation} (C$3$) is rated significantly higher than the static interaction (C$1$) for the \textit{anthropomorphism} ($U=0.5, p=0.004$), \textit{animacy} ($U=1.0, p=0.006$), and \textit{likeability}  ($U=2.0, p=0.011$) of the robot. Investigating the underlying dimensions for GODSPEED, C$3$ is rated significantly higher than C$1$ on \textit{conscious} ($U=2.5, p=0.007$), \textit{pleasant} ($U=2.0, p=0.005$) and \textit{nice} ($U=2.0, p=0.005$) dimensions while also being rated higher than C$2$ on the \textit{pleasant} ($U=5.5, p=0.004$) and \textit{nice} ($U=8.5, p=0.012$) dimensions. C$2$, on the other hand, is also rated higher than C$1$ on the \textit{conscious} ($U=5.0, p=0.008$) and \textit{responsive} ($U=5.5, p=0.011$) dimensions.

\begin{figure}  
    \centering
    \subfloat[RoSAS Scores.\label{fig:rosas}]{\includegraphics[width=0.45\textwidth]{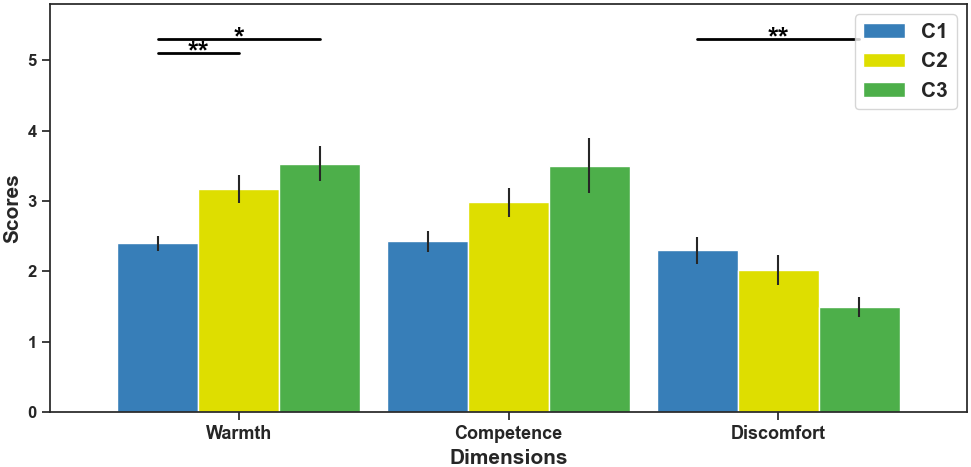}} \\\vspace{-4mm}
    \subfloat[Individual RoSAS Dimensions.\label{fig:rosas_dims}]{\includegraphics[width=\textwidth]{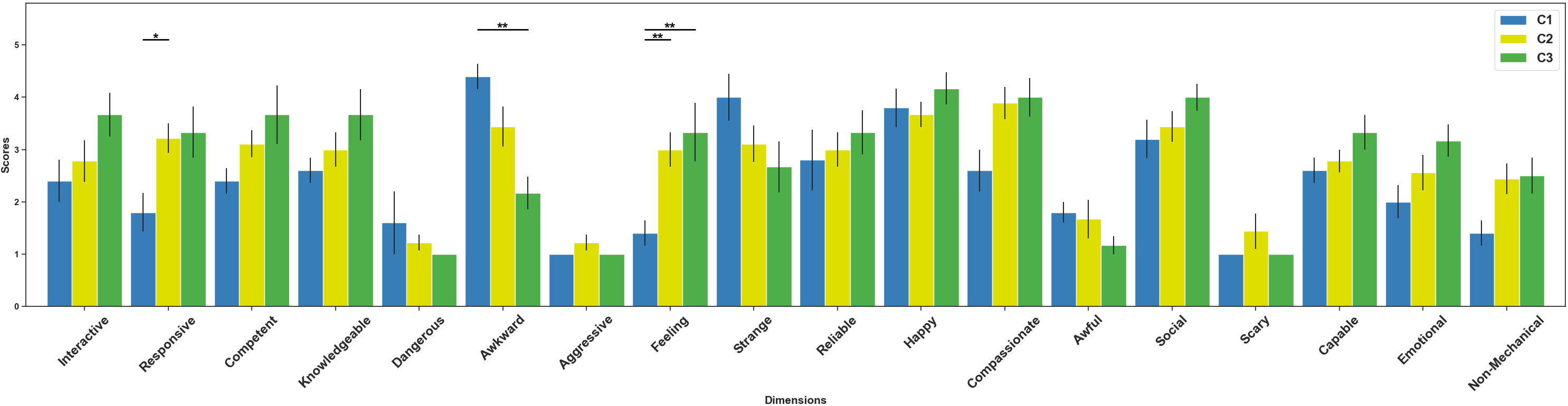}}\\
    \caption{RoSAS~\cite{carpinella2017robotic} Scores for C$1$, C$2$, and C$3$ conditions. $^*$ represents $p<0.05$ and $^{**}$ represents $p<0.01$.}
    \label{fig:rosas_combined} 
\end{figure}

\subsubsection{RoSAS}
Similar results are witnessed on the RoSAS evaluations as well, where C$2$ and C$3$ are rated better than C$1$ across all evaluations. Here too C$3$ performs the best being rated significantly higher than C$1$ for \textit{warmth} ratings ($U=2.0, p=0.010$) and significantly lower for \textit{discomfort} ratings ($U=29.0, p=0.006$). C$2$, on the other hand, is rated significantly higher for \textit{warmth} ratings ($U=, 4.5, p=0.009$) than C$1$. A deeper look into the underlying dimensions for RoSAS tells a similar story as the GODSPEED evaluations as C$3$ is consistently rated higher, on average, than both C$1$ and C$2$. It is rated significant \textit{less awkward} ($U=30.0, p=0.003$) and \textit{more feeling} ($U=2.0, p=0.008$) than C$1$. On the other hand, C$2$ is rated significantly \textit{more responsive} ($U=5.5, p=0.011$) and \textit{more feeling} ($U=5.5, p=0.008$)  than C$1$.
\begin{figure}
    \centering
    \includegraphics[width=0.45\textwidth]{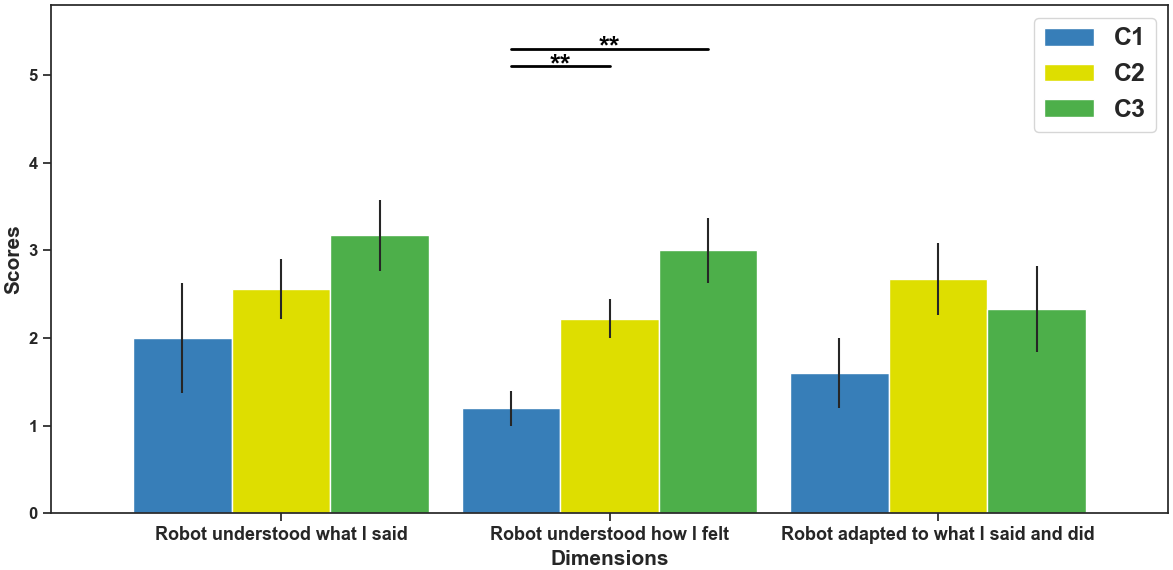}
    \caption{Customised Questions scores under C$1$, C$2$, and C$3$ conditions. $^{**}$ represents $p<0.01$.}
    \label{fig:custom}
\end{figure}

\subsubsection{Customised Questions}
\label{sec:customQs}

Customised questions are used to measure specific aspects of the interactions with Pepper such as its ability to understand what the user said, how they felt, and whether it adapted towards the participant. Pepper under C$3$ is rated higher than C$2$ and C$1$ across all dimensions with a significant difference witnessed in terms of understanding how the participants \textit{felt} ($U=1.0, p=0.005$), during the interactions. C$2$ is also rated significantly higher than C$1$ in understanding how the participants \textit{felt} ($U=5.5, p=0.009$). These results underscore the robot's ability, under C$2$ and C$3$ to be sensitive towards participants' affective behaviour during the interactions.

\section{Conclusion}
In this work, we design and implement a novel \ac{CL}-based framework on a physical robot to enable \textit{continual personalisation} capabilities, and undertake a \textit{proof-of-concept} study for wellbeing coaching. To the best of our knowledge, this is the first study evaluating how \textit{continual personalisation} capabilities in affective robots can improve participants' evaluation of them, improving their experiences with the robots. 

Our results from the \textit{proof-of-concept} user study show that a robot that adapts its behaviour by taking into account the affective behaviour of the participants (C$2$ and C$3$) is preferred, on average, over a static, non-adaptive (C$1$) one. Furthermore, embedding \textit{continual personalisation} capabilities in the robot (C$3$) significantly improves participants' impressions of the robot across several dimensions such as \textit{anthropomorphism, animacy, likeability, warmth} and \textit{comfort} and improves its perceived understanding of the participants' \textit{affective experience} during the interactions. For \ac{PP}-oriented wellbeing coaching, these are promising results as the expectation from such robots would be to offer positive experiences to the participants. Offering \textit{warm} and \textit{comfortable} interactions where the robot empathises with how the participants \textit{feel} allows for them to open up to the coaching provided, especially with the enhanced \textit{likeability} of the robot. This is evidenced in the reduced \textit{awkwardness} during the interactions with C$3$ being rated significantly higher in terms of being \textit{pleasant, nice} and \textit{feeling} towards the participants. Starting from a \textit{generalised} understanding of human facial affect, \ac{CL}-based personalisation towards each individual user ensures that the robot is able to accurately sense and understand their affective state during the interactions, enabling a dynamic and robust interaction with the participants, essential for wellbeing coaching. 

Even though adapting the interactions based on the participants' affective responses under both C$2$ and C$3$ offers improvements over making the robot always follow a static script, using \ac{CL} to personalise the robots' affect perception towards each participant makes these improvements significant. This is witnessed across evaluations where C$3$, on average, achieves the highest ratings. Thus, \textit{continual personalisation} may become an important capability for robots interacting with individuals, especially under wellbeing scenarios, where the robot is expected to be sensitive to their affective behaviour and adapt its behaviour accordingly. Such adaptation can have a positive impact on the interactions with the robot `empathising'~\cite{leite2013influence} with the participants and adapting the interaction based on their affective behaviour.

\subsection{Limitations and Future Work}

This \textit{proof-of-concept} evaluation constitutes preliminary results from a larger study aimed at investigating interactions with personalisable affective robots over repeated interactions. Here, we only investigate a one-off interaction to determine whether \ac{CL}-based personalisation improves the participants' interaction experience under wellbeing settings. Even though our evaluations provide promising results in favour of \textit{continual personalisation}, further analyses will focus on evaluating longitudinal interactions over multiple sessions to determine whether these effects hold under long-term \ac{HRI} settings. 

The screening of participants to make sure we were working with only a non-clinical population meant that several potential participants needed to be dropped from the user-study, leading to an imbalance in the gender distribution of the participants with more females participating in the study. Furthermore, technical issues during the experiments meant the data from some participants needed to be excluded from the analyses. This led to further imbalance in the distribution of the populations assigned randomly to each condition. Future work would also focus working on balancing these distributions for the participants for the longitudinal experiments.

Furthermore, we use a uni-modal affect perception for the robot evaluating only their facial expressions. This may not be sufficient as participants' verbal responses and their body gestures may also include important information about their affective responses during the interaction. Thus, future work should explore multi-modal affect perception to better evaluate the participants' affective responses. Furthermore, integrating more advanced \ac{NLP} capabilities will strengthen active listening capabilities for the robot, improving its contextual understanding of the interaction.

\section*{Acknowledgement}

\nl\textbf{Funding:} N.~Churamani is funded by the EPSRC under grant EP/R$513180$/$1$ (ref.~$2107412$). M.~Axelsson is funded by the Osk. Huttunen foundation and the EPSRC under grant EP/T$517847$/$1$. H.Gunes' work is supported by the EPSRC under grant ref. EP/R$030782$/$1$. A.Çaldır contributed to this work while undertaking a summer research study at the Department of Computer Science and Technology, University of Cambridge.

\nl\textbf{Data Access Statement:} Overall statistical analysis of research data underpinning this publication is contained in the manuscript. Additional raw data related to this publication cannot be openly released as the raw data contains videos and transcripts of the participants' interaction with the robot, which were impossible to anonymise.

\begin{acronym}
    \acro{AC}{Affective Computing}
    \acro{AI}{Artificial Intelligence}
    \acro{ASR}{Automatic Speech Recognition}
    \acro{AU}{Action Unit}
    \acro{BMU}{Best Matching Unit}
    \acro{CAAE}{Conditional Adversarial Auto-Encoder}
    \acro{CL}{Continual Learning}
    \acro{CLIFER}{Continual Learning with Imagination for Facial Expression Recognition}
    \acro{CNN}{Convolutional Neural Network}
    \acro{EDA}{Electrodermal Activity}
    \acro{FACS}{Facial Action Coding System}
    \acro{FER}{Facial Expression Recognition}
    \acro{FSM}{Finite-state Machine}
    \acro{FoI}{Frame-of-Interest}
    \acro{GAD7}{General Anxiety Disorder}
    \acro{GDM}{Growing Dual-Memory}
    \acro{GWR}{Growing When Required}
    \acro{HRI}{Human-Robot Interaction}
    \acro{MFCC}{Mel-Frequency Cepstral Coefficients}
    \acro{ML}{Machine Learning}
    \acro{MLP}{Multilayer Perceptron}
    \acro{NIC}{New Instances and Concepts}
    \acro{NLG}{Natural Language Generation}
    \acro{NLP}{Natural Language Processing}
    \acro{LSTM}{Long Short-Term Memory}
    \acro{PANAS}{Positive and Negative Affect Schedule}
    \acro{PD}{Participatory Design}
    \acro{PHQ9}{Participant Health Questionnaire}
    \acro{PP}{Positive Psychology}
    \acro{PPI}{Positive Psychology Intervention}
    \acro{RC}{Robotic Coach}
    \acro{ReLU}{Rectified Linear Unit}
    \acro{ROS}{Robot Operating System}
    \acro{RoSAS}{Robotic Social Attributes Scale}
    \acro{SAR}{Socially Assistive Robot}
    \acro{TA}{Thematic Analysis}
    \acro{TTS}{Text-to-Speech}
    \acro{WEMWBS}{Warwick-Edinburgh Mental Well-being Scale}
    \acro{WoZ}{Wizard of Oz}
\end{acronym}

\bibliographystyle{plainnat}
\bibliography{references.bib}

\begin{thebibliography}{42}
\providecommand{\natexlab}[1]{#1}
\providecommand{\url}[1]{\texttt{#1}}
\expandafter\ifx\csname urlstyle\endcsname\relax
  \providecommand{\doi}[1]{doi: #1}\else
  \providecommand{\doi}{doi: \begingroup \urlstyle{rm}\Url}\fi

\bibitem[Ayub and Wagner(2020)]{ayub2020b}
Ali Ayub and Alan~R. Wagner.
\newblock {What Am I Allowed to Do Here?: Online Learning of Context-Specific
  Norms by Pepper}.
\newblock In Alan~R. Wagner, David Feil-Seifer, Kerstin~S. Haring, Silvia
  Rossi, Thomas Williams, Hongsheng He, and Shuzhi Sam~Ge, editors,
  \emph{Social Robotics ICSR 2020}, pages 220--231. Springer, 2020.
\newblock ISBN 978-3-030-62056-1.

\bibitem[Barros et~al.(2019)Barros, Parisi, and Wermter]{pmlrbarros19a}
Pablo Barros, German Parisi, and Stefan Wermter.
\newblock {A Personalized Affective Memory Model for Improving Emotion
  Recognition}.
\newblock In Kamalika Chaudhuri and Ruslan Salakhutdinov, editors,
  \emph{{Proceedings of the 36th International Conference on Machine
  Learning}}, volume~97 of \emph{Proceedings of Machine Learning Research},
  pages 485--494. PMLR, Jun 2019.

\bibitem[Barros et~al.(2020{\natexlab{a}})Barros, Churamani, and
  Sciutti]{BCS2020FaceSN}
Pablo Barros, Nikhil Churamani, and Alessandra Sciutti.
\newblock The {FaceChannel}: A fast and furious deep neural network for facial
  expression recognition.
\newblock \emph{{SN} Computer Science}, 1\penalty0 (6), October
  2020{\natexlab{a}}.

\bibitem[Barros et~al.(2020{\natexlab{b}})Barros, Churamani, and
  Sciutti]{barros2020facechannel}
Pablo Barros, Nikhil Churamani, and Alessandra Sciutti.
\newblock {The FaceChannel: A Light-weight Deep Neural Network for Facial
  Expression Recognition}.
\newblock In \emph{Proceedings of the 15th International Conference on
  Automatic Face and Gesture Recognition (FG)}, pages 652--656,
  2020{\natexlab{b}}.

\bibitem[Bartneck et~al.(2009)Bartneck, Kuli{\'c}, Croft, and
  Zoghbi]{bartneck2009measurement}
Christoph Bartneck, Dana Kuli{\'c}, Elizabeth Croft, and Susana Zoghbi.
\newblock Measurement instruments for the anthropomorphism, animacy,
  likeability, perceived intelligence, and perceived safety of robots.
\newblock \emph{International Journal of Social Robotics}, 1\penalty0
  (1):\penalty0 71--81, 2009.

\bibitem[Bodala et~al.(2021)Bodala, Churamani, and
  Gunes]{Bodala2021Teleoperated}
Indu~P. Bodala, Nikhil Churamani, and Hatice Gunes.
\newblock {Teleoperated Robot Coaching for Mindfulness Training: A Longitudinal
  Study}.
\newblock In \emph{30th IEEE International Conference on Robot and Human
  Interactive Communication (RO-MAN)}, pages 939--944, 2021.

\bibitem[Carpinella et~al.(2017)Carpinella, Wyman, Perez, and
  Stroessner]{carpinella2017robotic}
Colleen~M Carpinella, Alisa~B Wyman, Michael~A Perez, and Steven~J Stroessner.
\newblock The robotic social attributes scale (rosas) development and
  validation.
\newblock In \emph{Proceedings of the ACM/IEEE International Conference on
  Human-Robot Interaction}, pages 254--262, 2017.

\bibitem[{Chu} et~al.(2017){Chu}, l.~{Torre}, and {Cohn}]{Chu2017Selective}
W.~{Chu}, F.~D. l.~{Torre}, and J.~F. {Cohn}.
\newblock {Selective Transfer Machine for Personalized Facial Expression
  Analysis}.
\newblock \emph{{IEEE Transactions on Pattern Analysis and Machine
  Intelligence}}, 39 (3):\penalty0 529--545, 2017.
\newblock ISSN 0162-8828.

\bibitem[Churamani(2020)]{Churamani2020CLAC}
Nikhil Churamani.
\newblock {Continual Learning for Affective Computing}.
\newblock In \emph{Doctoral Consortium Proceedings of the 15th International
  Conference on Automatic Face and Gesture Recognition (FG)}, 2020.
\newblock {arXiv:2006.06113}.

\bibitem[Churamani and Gunes(2020)]{Churamani2020CLIFER}
Nikhil Churamani and Hatice Gunes.
\newblock {CLIFER: Continual Learning with Imagination for Facial Expression
  Recognition}.
\newblock In \emph{Proceedings of the 15th IEEE International Conference on
  Automatic Face and Gesture Recognition (FG)}, pages 322--328. IEEE, 2020.

\bibitem[Churamani et~al.(2017)Churamani, Anton, Br\"{u}gger, Flie\ss~wasser,
  Hummel, Mayer, Mustafa, Ng, Nguyen, Nguyen, Soll, Springenberg, Griffiths,
  Heinrich, Navarro-Guerrero, Strahl, Twiefel, Weber, and
  Wermter]{Churamani2017TheImpact}
Nikhil Churamani, Paul Anton, Marc Br\"{u}gger, Erik Flie\ss~wasser, Thomas
  Hummel, Julius Mayer, Waleed Mustafa, Hwei~Geok Ng, Thi Linh~Chi Nguyen, Quan
  Nguyen, Marcus Soll, Sebastian Springenberg, Sascha Griffiths, Stefan
  Heinrich, Nicol\'{a}s Navarro-Guerrero, Erik Strahl, Johannes Twiefel,
  Cornelius Weber, and Stefan Wermter.
\newblock The impact of personalisation on human-robot interaction in learning
  scenarios.
\newblock In \emph{Proceedings of the 5th International Conference on Human
  Agent Interaction}, pages 171--180. ACM, 2017.
\newblock ISBN 978-1-4503-5113-3.

\bibitem[Churamani et~al.(2020)Churamani, Kalkan, and
  Gunes]{Churamani2020CL4AR}
Nikhil Churamani, Sinan Kalkan, and Hatice Gunes.
\newblock Continual learning for affective robotics: Why, what and how?
\newblock In \emph{29th IEEE International Conference on Robot and Human
  Interactive Communication (RO-MAN)}, pages 425--431, 2020.

\bibitem[Churamani et~al.(2022)Churamani, Barros, Gunes, and
  Wermter]{Churamani2022Affect}
Nikhil Churamani, Pablo Barros, Hatice Gunes, and Stefan Wermter.
\newblock Affect-driven learning of robot behaviour for collaborative
  human-robot interactions.
\newblock \emph{Frontiers in Robotics and AI}, 9, 2022.
\newblock ISSN 2296-9144.

\bibitem[Cunha et~al.(2019)Cunha, Pellanda, and Reppold]{cunha2019positive}
L{\'u}zie~Fofonka Cunha, Lucia~Campos Pellanda, and Caroline~Tozzi Reppold.
\newblock Positive psychology and gratitude interventions: A randomized
  clinical trial.
\newblock \emph{Frontiers in Psychology}, 10, 2019.

\bibitem[Dautenhahn(2004)]{Dautenhahn2004Robots}
Kerstin Dautenhahn.
\newblock Robots we like to live with?! - a developmental perspective on a
  personalized, life-long robot companion.
\newblock In \emph{IEEE International Workshop on Robot and Human Interactive
  Communication (RO-MAN)}, pages 17--22, Sept 2004.

\bibitem[Dziergwa et~al.(2018)Dziergwa, Kaczmarek, Kaczmarek, Kedzierski, and
  Wadas-Szyd{\l}owska]{Dziergwa2018}
Micha{\l} Dziergwa, Mirela Kaczmarek, Pawe{\l} Kaczmarek, Jan Kedzierski, and
  Karolina Wadas-Szyd{\l}owska.
\newblock {Long-Term Cohabitation with a Social Robot: A Case Study of the
  Influence of Human Attachment Patterns}.
\newblock \emph{International Journal of Social Robotics}, 10\penalty0
  (1):\penalty0 163--176, Jan 2018.
\newblock ISSN 1875-4805.

\bibitem[Ficocelli et~al.(2016)Ficocelli, Terao, and
  Nejat]{Ficocelli2016Promoting}
Maurizio Ficocelli, Junichi Terao, and Goldie Nejat.
\newblock {Promoting Interactions Between Humans and Robots Using Robotic
  Emotional Behavior}.
\newblock \emph{IEEE Transactions on Cybernetics}, 46\penalty0 (12):\penalty0
  2911--2923, 2016.

\bibitem[Griffiths et~al.(2018)Griffiths, Alpay, Sutherland, Kerzel, Eppe,
  Strahl, and Wermter]{GASKESW18}
Sascha Griffiths, Tayfun Alpay, Alexander Sutherland, Matthias Kerzel, Manfred
  Eppe, Erik Strahl, and Stefan Wermter.
\newblock Exercise with social robots: Companion or coach?
\newblock In \emph{Proceedings of Workshop on Personal Robots for Exercising
  and Coaching at the ACM/IEEE Interanational Conference on Human-Robot
  Interaction}. ACM, Mar 2018.

\bibitem[Gunes et~al.(2011)Gunes, Schuller, Pantic, and
  Cowie]{Gunes2011Emotion}
Hatice Gunes, Bj\"orn Schuller, Maja Pantic, and Roddy Cowie.
\newblock Emotion representation, analysis and synthesis in continuous space: A
  survey.
\newblock In \emph{2011 IEEE International Conference on Automatic Face and
  Gesture Recognition (FG)}, pages 827--834, 2011.

\bibitem[Hemminghaus and Kopp(2017)]{Hemminghaus2017Adaptive}
Jacqueline Hemminghaus and Stefan Kopp.
\newblock Towards adaptive social behavior generation for assistive robots
  using reinforcement learning.
\newblock In \emph{Proceedings of the 2017 ACM/IEEE International Conference on
  Human-Robot Interaction}, page 332–340, 2017.
\newblock ISBN 9781450343367.

\bibitem[Kirby et~al.(2010)Kirby, Forlizzi, and Simmons]{KIRBY2010Affective}
Rachel Kirby, Jodi Forlizzi, and Reid Simmons.
\newblock Affective social robots.
\newblock \emph{Robotics and Autonomous Systems}, 58\penalty0 (3):\penalty0
  322--332, 2010.

\bibitem[Kroenke et~al.(2001)Kroenke, Spitzer, and Williams]{kroenke2001phq}
Kurt Kroenke, Robert~L Spitzer, and Janet~BW Williams.
\newblock {The PHQ-9: validity of a brief depression severity measure}.
\newblock \emph{Journal of General Internal Medicine}, 16\penalty0
  (9):\penalty0 606--613, 2001.

\bibitem[Kruskal and Wallis(1952)]{Kruskal1952KW}
William~H. Kruskal and W.~Allen Wallis.
\newblock Use of ranks in one-criterion variance analysis.
\newblock \emph{Journal of the American Statistical Association}, 47\penalty0
  (260):\penalty0 583--621, 1952.
\newblock ISSN 01621459.

\bibitem[Leite et~al.(2013{\natexlab{a}})Leite, Martinho, and
  Paiva]{Leite2013SocialRF}
Iolanda Leite, Carlos Martinho, and Ana Paiva.
\newblock Social robots for long-term interaction: A survey.
\newblock \emph{International Journal of Social Robotics}, 5:\penalty0
  291--308, 2013{\natexlab{a}}.

\bibitem[Leite et~al.(2013{\natexlab{b}})Leite, Pereira, Mascarenhas, Martinho,
  Prada, and Paiva]{leite2013influence}
Iolanda Leite, Andr{\'e} Pereira, Samuel Mascarenhas, Carlos Martinho, Rui
  Prada, and Ana Paiva.
\newblock The influence of empathy in human--robot relations.
\newblock \emph{International Journal of Human-Computer Studies}, 71\penalty0
  (3):\penalty0 250--260, 2013{\natexlab{b}}.

\bibitem[Lesort et~al.(2020)Lesort, Lomonaco, Stoian, Maltoni, Filliat, and
  Díaz-Rodríguez]{LESORT2020CL4R}
Timothée Lesort, Vincenzo Lomonaco, Andrei Stoian, Davide Maltoni, David
  Filliat, and Natalia Díaz-Rodríguez.
\newblock {Continual learning for robotics: Definition, framework, learning
  strategies, opportunities and challenges}.
\newblock \emph{Information Fusion}, 58:\penalty0 52--68, 2020.
\newblock ISSN 1566-2535.

\bibitem[{Li} and {Deng}(2020)]{Li2020Deep}
S.~{Li} and W.~{Deng}.
\newblock Deep facial expression recognition: A survey.
\newblock \emph{IEEE Transactions on Affective Computing}, March 2020.

\bibitem[{Lindt} et~al.(2019){Lindt}, {Barros}, {Siqueira}, and
  {Wermter}]{Lindt2019FG}
A.~{Lindt}, P.~{Barros}, H.~{Siqueira}, and S.~{Wermter}.
\newblock {Facial Expression Editing with Continuous Emotion Labels}.
\newblock In \emph{Proceedings of the 14th IEEE International Conference on
  Automatic Face and Gesture Recognition (FG)}, pages 1--8. IEEE, 2019.

\bibitem[Lohan et~al.(2016)Lohan, Lehmann, Dondrup, Broz, and Kose]{Lohan2016}
Katrin~Solveig Lohan, Hagen Lehmann, Christian Dondrup, Frank Broz, and Hatice
  Kose.
\newblock \emph{Enriching the Human-Robot Interaction Loop with Natural,
  Semantic, and Symbolic Gestures}, pages 1--21.
\newblock Springer Netherlands, Dordrecht, 2016.
\newblock ISBN 978-94-007-7194-9.

\bibitem[Lopez et~al.(2018)Lopez, Pedrotti, and Snyder]{lopez2018positive}
Shane~J Lopez, Jennifer~Teramoto Pedrotti, and Charles~Richard Snyder.
\newblock \emph{Positive psychology: The scientific and practical explorations
  of human strengths}.
\newblock Sage publications, 2018.

\bibitem[Mann and Whitney(1947)]{Mann1947}
H~B Mann and D~R Whitney.
\newblock {On a test of whether one of two random variables is stochastically
  larger than the other.}
\newblock \emph{The Annals of Mathematical Statistics}, 18:\penalty0 50--60,
  1947.

\bibitem[McQuillin et~al.(2022)McQuillin, Churamani, and
  Gunes]{McQuillin2022RoboWaiter}
Emily McQuillin, Nikhil Churamani, and Hatice Gunes.
\newblock {Learning Socially Appropriate Robo-Waiter Behaviours through
  Real-Time User Feedback}.
\newblock In \emph{Proceedings of the 2022 ACM/IEEE International Conference on
  Human-Robot Interaction (HRI)}, pages 541--–550. IEEE Press, 2022.

\bibitem[Miller and Feil-Seifer(2016)]{Miller2016}
Blanca Miller and David Feil-Seifer.
\newblock \emph{Embodiment, Situatedness, and Morphology for Humanoid Robots
  Interacting with People}, pages 1--23.
\newblock Springer, 2016.
\newblock ISBN 978-94-007-7194-9.

\bibitem[Noroozi et~al.(2021)Noroozi, Corneanu, Kaminska, Sapinski, Escalera,
  and Anbarjafari]{Noroozi2021Survey}
F.~Noroozi, C.~Corneanu, D.~Kaminska, T.~Sapinski, S.~Escalera, and
  G.~Anbarjafari.
\newblock Survey on emotional body gesture recognition.
\newblock \emph{IEEE Transactions on Affective Computing}, 12\penalty0
  (02):\penalty0 505--523, apr 2021.
\newblock ISSN 1949-3045.

\bibitem[Parisi et~al.(2018)Parisi, Tani, Weber, and Wermter]{Parisi2018a}
German~I. Parisi, Jun Tani, Cornelius Weber, and Stefan Wermter.
\newblock {Lifelong Learning of Spatiotemporal Representations With Dual-Memory
  Recurrent Self-Organization}.
\newblock \emph{{Frontiers in Neurorobotics}}, 12:\penalty0 78, October 2018.
\newblock ISSN 1662-5218.

\bibitem[Parisi et~al.(2019)Parisi, Kemker, Part, Kanan, and
  Wermter]{Parisi2018b}
G.I. Parisi, R.~Kemker, J.L. Part, C.~Kanan, and S.~Wermter.
\newblock {Continual Lifelong Learning with Neural Networks: A review}.
\newblock \emph{Neural Networks}, 113:\penalty0 54--71, 2019.

\bibitem[Rudovic et~al.(2018)Rudovic, Lee, Dai, Schuller, and
  Picard]{Rudoviceaao6760}
Ognjen Rudovic, Jaeryoung Lee, Miles Dai, Bj{\"o}rn Schuller, and Rosalind~W.
  Picard.
\newblock {Personalized machine learning for robot perception of affect and
  engagement in autism therapy}.
\newblock \emph{Science Robotics}, 3\penalty0 (19), 2018.

\bibitem[Russell(1980)]{russell1980circumplex}
James~A Russell.
\newblock A circumplex model of affect.
\newblock \emph{Journal of Personality and Social Psychology}, 39\penalty0
  (6):\penalty0 1161, 1980.

\bibitem[Schuller(2018)]{Schuller2018SER}
Bj\"{o}rn~W. Schuller.
\newblock Speech emotion recognition: Two decades in a nutshell, benchmarks,
  and ongoing trends.
\newblock \emph{Commun. ACM}, 61\penalty0 (5):\penalty0 90--99, April 2018.
\newblock ISSN 0001-0782.

\bibitem[Spitzer et~al.(2006)Spitzer, Kroenke, Williams, and
  L{\"o}we]{spitzer2006brief}
Robert~L Spitzer, Kurt Kroenke, Janet~BW Williams, and Bernd L{\"o}we.
\newblock {A brief measure for assessing generalized anxiety disorder: the
  GAD-7}.
\newblock \emph{Archives of Internal Medicine}, 166\penalty0 (10):\penalty0
  1092--1097, 2006.

\bibitem[Voghoei et~al.(2019)Voghoei, Tonekaboni, Wallace, and
  Arabnia]{dlAtTheEdge}
Sahar Voghoei, Navid~Hashemi Tonekaboni, Jason~G. Wallace, and Hamid~R.
  Arabnia.
\newblock Deep learning at the edge.
\newblock \emph{CoRR}, abs/1910.10231, 2019.

\bibitem[Wullenkord and Eyssel(2019)]{Wullenkord2019}
Ricarda Wullenkord and Friederike Eyssel.
\newblock Imagine how to behave: the influence of imagined contact on
  human{\textendash}robot interaction.
\newblock \emph{Philosophical Transactions of the Royal Society B: Biological
  Sciences}, 374\penalty0 (1771):\penalty0 20180038, April 2019.

\end{thebibliography}

\end{document}